\documentclass[a4paper,11pt]{article}
\usepackage{epsfig}
\usepackage{amssymb}
\usepackage{amsmath}

\begin{document}

\title{Temperature Inversion Symmetry in Gauge-Higgs Unification Models}
\author{V.K.Oikonomou\thanks{%
voiko@physics.auth.gr}\\
Dept. of Theoretical Physics Aristotle University of Thessaloniki,\\
Thessaloniki 541 24 Greece}
\maketitle

\begin{abstract}
The temperature inversion symmetry $R\rightarrow \frac{1}{T}$ is
studied for the finite temperature effective potential of the
$N=1$, $d=5$, supersymmetric $SU(3)_{c}{\times}SU(3)_{w}$ model,
on the orbifold $S^{1}/Z_{2}$. For the value of the Wilson line
parameter $\alpha=1$ ($SU(2)_{L}$ breaks to a $U'(1)$), it is
found that the effective potential contains a symmetric part and
an anti-symmetric part under $\xi\rightarrow \frac{1}{\xi}$, with,
$\xi=RT$. When $\alpha=0$ (for which, $SU(2)_{L}$ remains
unbroken) it is found that the only contribution to the effective
potential that spoils the temperature inversion symmetry comes
from the fermions in the fundamental representation of the gauge
group, with $(+,+)$ or $(-,-)$, $Z_{2}$ parities. This is
interesting since it implies that the bulk effective potential
corresponding to models with orbifold fixed point localized
fundamental fermions (and with no bulk fundamental fermions) has
the temperature inversion symmetry.
\end{abstract}

\bigskip
\bigskip
\section*{Introduction}

The importance of dualities in quantum field theories is
undoubtable, especially for effective theories that their
predictions lie beyond the perturbative limits or the current
experimental bounds. The temperature inversion symmetry (expressed
by the transformation $R\rightarrow \frac{1}{T}$), if it holds,
connects the zero temperature vacuum energy of a field theory,
with the Boltzmann free energy. Thus the study of such a symmetry
in field theoretic models is of great importance since it connects
two conceptually different limits of the same theory.

As stated above, temperature inversion symmetry stands for the
transformation $R\rightarrow \frac{1}{T}$. The name temperature
inversion symmetry \cite{wotzasek} is misleading but it is
traditionally used to describe systems that, under the
transformation $R\rightarrow \frac{1}{T}$, their Boltzmann free
energy (high temperature limit) is equal to the Casimir vacuum
energy at zero temperature. For example consider a $d=4$
supersymmetric non-interacting ensemble of periodic fermions and
anti-periodic bosons. The free energy of this system at high
temperature is equal to:
\begin{equation}
F=-T^{4}\pi ^{-2}(\frac{\pi ^{4}}{45}+\frac{7\pi ^{4}}{360}),
\label{4dim}
\end{equation}
while the Casimir energy of the ensemble is:
\begin{equation}
E_{o}=\frac{1}{R^{4}}\pi ^{-2}(\frac{\pi ^{4}}{45}+\frac{7\pi
^{4}}{360}). \label{4dimcasimir}
\end{equation}
Using the transformation $R\rightarrow \frac{1}{T}$ we can see
that $F=-E_o$, because the system is anti-symmetric under the
temperature inversion symmetry.

There is an intriguing similarity of this symmetry with the
$T$-duality \cite{strings} of the closed bosonic string. The total
squared mass of the bosonic string is:
\begin{equation}\label{bosonicstring}
m^2=\frac{n^2}{R^2}+\frac{w^2
R^2}{{\alpha^{\prime}}^2}+\frac{2}{\alpha^{\prime}}(N+\tilde{N}-2)
\end{equation}
with ${\alpha^{\prime}}^{1/2}$ the self dual radius. We can see
that the total mass square is invariant under the transformation
$R\rightarrow \frac{\alpha^{\prime}}{R}$ and $n\rightarrow w$.
This is known as $T$-duality for the bosonic string and it
connects the $R\rightarrow 0$ with the $R\rightarrow \infty$
limits of the theory (actually the two limits describe the same
theory). This is an intrinsic feature of string theory. The total
mass square contains the Kaluza-Klein excitations
$\frac{n^2}{R^2}$ and the winding modes $\frac{w^2
R^2}{{\alpha^{\prime}}^2}$. The winding modes are absent from
field theories and that is why $T$-duality is not a symmetry of
field theory models (and so the $R\rightarrow \infty$ and
$R\rightarrow 0$ limits are disconnected).

However the Kaluza-Klein mass at finite temperature of a periodic
boson in field theory (for a compact dimension of radius $R$) is:
\begin{equation}\label{fieldtheor}
m^2_{n,m}=\frac{4\pi^2 n^2}{R^2}+4\pi^2 m^2T^2
\end{equation}
The winding modes are absent but there are ''thermal winding
modes'' $\sim\nobreak m^2T^2$ (if we are allowed to use this
expression). In a way temperature inversion symmetry can serve as
the corresponding $T$-duality of field theory. Indeed the
transformation $R\rightarrow \frac{1}{T}$ leaves the thermal
Kaluza-Klein mass invariant.

It worths mentioning another similarity. The closed bosonic string
free energy transforms under the thermal duality \cite{dienes}
transformation $T\rightarrow T_c^2/T$ (closely related to the
$T$-duality) as:
\begin{equation}\label{thermaldualy}
\mathcal{F}\Big{(}\frac{T_{c}^{2}}{T}\Big{)}=\Big{(}\frac{T_{c}}{T}\Big{)}^{2}\mathcal{F}(T)
\end{equation}
with
\begin{equation}\label{selfdual}
T_{c}=\frac{M_{str}}{2\pi}=\frac{1}{2\pi{\alpha^{\prime}}^{1/2}}
\end{equation}
the self dual temperature (Hagedorn) and $M_{str}$, the string
scale. Due to the thermal duality the high temperature limit (well
above the Hagedorn temperature) is related to the low temperature
limit (well below the Hagedorn temperature). In the field theory
case, as far as the temperature inversion symmetry is concerned,
no similar situation occurs, which means that there is now way to
relate the high and low temperature limit of the same theory.
However there exists a similarity in the context of the
temperature inversion symmetry. In field theory, the scaled
bosonic free energy $f(\xi)$, with $\xi=RT$, is covariant under
$\xi\rightarrow \frac{1}{\xi}$ and obeys the relation
\begin{equation}
f(\xi )=\xi ^{4}f(\frac{1}{\xi })
\end{equation}
According to the above considerations we could say that the
temperature inversion symmetry is the corresponding combination of
the string theory's $T$-{\nobreak}duality and thermal duality for
field theory models with compact dimensions at finite temperature.

In general, when studying four dimensional models with one compact
dimension, the scaled finite temperature effective potential of a
boson is symmetric under temperature inversion symmetry, while the
effective potential of a periodic fermion and anti-periodic boson
(periodicity referring to the compact dimension boundary
condition) is antisymmetric. Moreover when bosons are periodic and
fermions antiperiodic, the vacuum energy is symmetric under
$R\rightarrow \frac{1}{T}$ \cite{wotzasek}.

It is clear that the boundary conditions used in circle
compactifications affect the transformations under $R\rightarrow
\frac{1}{T}$, making the ensemble either symmetric or
antisymmetric.

In this paper we extend the study of temperature inversion
symmetry to five dimensional orbifold models. We shall consider
how orbifold boundary conditions modify the transformation of the
effective potential under $R\rightarrow \frac{1}{T}$.

Field theoretic orbifold compactifications are very useful from
many theoretical points of view, for example the problem of
fermions chirality in higher dimensional field theories, which can
be solved through field theoretic orbifold compactifications
\cite{qui1,qui2,extra2} (originally used in string field theory
compactifications). On the orbifold compactification setup,
fermions can be localized on $4d$ dimensional hypersurfaces called
branes, which are curvature singular fixed points of the orbifold.
Thus chirality can be achieved. Also the use of orbifolds offers
many theoretical uses, such as, gauge symmetry breaking
\cite{russell1,russell2,hoso2} and higher dimensional
supersymmetry breaking
\cite{kawamura,extra2,extra1,barbieri,russell1}. Many alternative
models in $d=5$ and $d=6$ dimensions have been introduced mainly
using $S^{1}/Z_{2}$ and $S^{1}/Z_{2}\times Z'_{2}$ orbifold
structure for the extra dimension. The primary objective of all
extra dimensional models is to mimic at low energies, the Standard
Model, or the MSSM. One promising class of models is the so called
gauge-Higgs unification models \cite{inigauge,gaugehiggs}, where
the Higgs field is identified with the higher dimensional
component of a gauge field. In this paper an $S^{1}/Z_{2}$
orbifold compactification gauge-Higgs modelling shall be used.
Usually in these models one starts with a higher dimensional (five
or six dimensional) gauge theory (supersymmetric or not) of some
simple gauge group ({\it{i.e.}} that cannot be written as a
product of groups and thus can have only one coupling) and breaks
the gauge group through the orbifold boundary conditions. Further
breaking of the gauge symmetry can be achieved radiatively through
the Hosotani mechanism \cite{hoso1} (Wilson line breaking) since
the vacuum expectation value of the Higgs field is proportional to
the Wilson line phase.

In this paper we shall study the bulk effective potential for the
$N=1$, $d=5$, supersymmetric $SU(3)_{c}{\times}SU(3)_{w}$
gauge-Higgs unification model at finite temperature
\cite{takenaga}. For this model we shall use the $S^{1}/Z_{2}$
orbifold compactification boundary conditions for the fields and
go over the temperature inversion symmetry \cite{wotzasek}
$R\rightarrow \frac{1}{T}$ for the bulk effective potential. The
(Wilson line) minima of the effective potential $\alpha=0,1$ and
Scherk-Schwarz phases for the s-parteners shall be used.
Furthermore, the particle content shall be that of a $N=1$, $d=5$
vector multiplet, which in $d=4$ dimensions is equivalent to an
$N=2$ supersymmetry vector multiplet \cite{arkani,gaume} (one
vector and one chiral $d=4$ multiplet). Also bulk matter fields,
with various flavor numbers and group representations shall be
added, in order to study their effect on temperature inversion
symmetry of the corresponding effective potential (for a very nice
calculation method of the effective potential see \cite{hab1}).

In the next section a brief review of the orbifold
compactification procedure shall be given.

\bigskip
\section*{Brief Review of Orbifolding}
Orbifolding is a string inspired technique that was originally
used in $TeV$ physics in order to obtain chiral fermions from a
higher dimensional vector-like theory \cite{qui1,extra1,extra2}.
Later uses include supersymmetry breaking and gauge symmetry
breaking on the orbifold fixed points
\cite{russell1,russell2,kawamura,barbieri}. Orbifolds can be
constructed by acting on a compact manifold $C$ non freely, with a
discrete group $H$. Under the transformation of the group,
represented by $\zeta_{h}$, the point $y$ of the compact manifold
$C$ is identified with $\zeta_{h}(y)$ that is,
\begin{equation}
y\sim{\zeta_{h}}(y)\label{equiv}.
\end{equation}
The lagrangian of the field theory must be invariant under the
equivalence relation (\ref{equiv}). If $\phi(x,y)$ is a field
representing all fields and $y$ the coordinate of the compact
space to be orbifolded, then:
\begin{equation}
L(\phi{(x,y))}=L(\phi{(x,\zeta_{h}(y))},
\end{equation}
which in terms of the fields becomes,
\begin{equation}
\phi{(x,\zeta_{h}(y))}=Z_{h}\phi{(x,y)}.
\end{equation}
$Z_{h}$ has to be a symmetry of the lagrangian and the various
fields must transform in such way that the above symmetry holds.
The non free action of $H$ on $C$ means that the transformation
$\zeta_{h}$ has fixed points. Thus the resulting quotient space
$O\equiv{C/H}$ is not a manifold but a singular space at the fixed
points of the transformation called orbifold. It has to be
mentioned that if $\phi(x,y)$ is a representation of some group
then $Z_{h}$ can be a non trivial representation of the discrete
group $H$. This way, gauge symmetry can be broken between the
components of the gauge multiplet. We describe below the example
of $S^{1}/Z_{2}$ orbifold which shall be used in this paper.

\subsection*{Scherk-Schwarz breaking}

In this subsection the Scherk-Schwarz mechanism \cite{scherk} is
described and the conventions of this section shall be implied in
the following. The Scherk-Schwarz supersymmetry breaking mechanism
is based on imposing different boundary conditions between
fermions and bosons under the transformation $y\rightarrow
y+2{\pi}R$, that is \cite{qui2}:
\begin{equation}
\Phi(x,y+2\pi R)=e^{{i 2\pi}{\beta}}\Phi(x,y).
\end{equation}
Due to the above boundary condition, the fields have KK expansions
as
\begin{equation}
\Phi(x,y)=\sum_{n=-\infty}^{\infty}e^{\frac{i
y{(n+\beta)}}{R}}{\Phi}^{n}(x),
\end{equation}
or
\begin{equation}
\Phi(x,y)=\sum_{n=-\infty}^{\infty}\cos
{\big{(}}\frac{(n+\beta)y}{R}{\big{)}}{\Phi}^{n}(x),
\end{equation}
and
\begin{equation}
\Phi(x,y)=\sum_{n=-\infty}^{\infty}\sin
{\big{(}}\frac{(n+\beta)y}{R}{\big{)}}{\Phi}^{n}(x),
\end{equation}
 for the orbifold case, in general.

\subsection*{The $S^{1}/Z_{2}$ orbifold}

One of the most frequently used one dimensional field theoretic
orbidold, is $S^{1}/Z_{2}$. It is constructed by identifying a
point $y$ on $S^{1}$ with $-y$, that is, a $Z_{2}$ equivalence
relation \cite{kim}:
\begin{equation}
Z_{2}: y\rightarrow{-y},
\end{equation}
with $S^{1}$ coordinatized as $-\pi{R}<y\leq{\pi{R}}$ and $R$ the
circle radius. The $Z_{2}$ action has two fixed points $y=0$ and
$y=\pi{R}$ which in our world terminology, in terms of
$M^{4}\times{S^{1}/Z_{2}}$ space-time, are called branes. Under
the $Z_{2}$ identification, the resulting space is an orbifold
coordinatized as $0\leq y\leq{\pi{R}}$. A quantum field (that may
be a representation of a gauge group) transforms under the $Z_{2}$
action as:
\begin{equation}
\phi{(x,-y)}=Z\phi{(x,y)}.
\end{equation}
It can be easily seen that $Z^{2}=I$. So in the field
representation space, $Z$ can be diagonalized with eigenvalues
$\pm 1$. This means that $Z$ can be $I$, $-I$, or a mixed diagonal
matrix $P$ with eigenvalues $\pm 1$. The later can be used to
break gauge symmetry (see below).

Orbifolds will be used later on to break gauge symmetry, so a
simple example follows here, to see how this works out. Consider a
pure $5d$ $SU(3)$ model, with action:
\begin{equation}
S=\int \mathrm{d}x^{4}\mathrm{d}y(-\frac{1}{4}
F^{MN}F_{MN})\label{A},
\end{equation}
with $M,N=1,2,3,4,5$. In order to break $SU(3)$ we choose the
following boundary condition for the $SU(3)$ Lie algebra valued
gauge field connection $A_{M}$:
\begin{equation}
A_{M}(x,y)=\Lambda ^{N}_{M}PA_{N}(x,-y)P^{\dag}\label{B}
\end{equation}
where
\begin{equation}
\Lambda{=\left(%
\begin{array}{ccccc}
  1 & 0 & 0 & 0 & 0 \\
  0 & 1 & 0 & 0 & 0 \\
  0 & 0 & 1 & 0 & 0 \\
  0 & 0 & 0 & 1 & 0 \\
  0 & 0 & 0 & 0 & -1 \\
\end{array}%
\right)},
\end{equation}
which is determined from the requirement of $5d$ Lorentz
invariance \cite{arkani} and
\begin{equation}
\left(%
\begin{array}{ccc}
  -1 & 0 & 0 \\
  0 & -1 & 0 \\
  0 & 0 & 1 \\
\end{array}%
\right)\label{C}.
\end{equation}
Under this specific representation $P$ of $Z_{2}$ (note that
$P^2=1$), $A^{1,2,3}_{\mu}$ and $A^8_{\mu}$ are even, while,
$A^{4,5,6,7}_{\mu}$ are odd. Thus on the orbifold fixed points,
{\it{i.e.}} in $4d$, $SU(3)$ is broken to $SU(2)\times U(1)$,
since the harmonic expansion of each of $A^{4,5,6,7}_{\mu}$ does
not have a zero mode. So the well below the compactification scale
particle spectrum is arranged to $SU(2)\times U(1)$ multiplets.
Note that the original gauge symmetry is reduced to the
centralizer of $P$ in $SU(3)$ (or equivalently, only the field
components with $[P,A^{\alpha}_{\mu}]\neq 0$ break the gauge
symmetry). It must be mentioned that there exist another one
discrete symmetry $Z'_{2}$ on top the one described above, when
the extra dimension number is one. The extra orbifolding can lead
to supersymmetry breaking and gauge symmetry breaking in some
models \cite{barbieri,inigauge,russell1}.

\bigskip

\section*{Gauge-Higgs Unification modelling on $S^{1}/Z_{2}$}

Consider for simplicity a $5d$ $SU(N)$ gauge theory on
$S^{1}/Z_{2}\times M^{4}$ (the notation of \cite{hoso2} is used).
Gauge fields are bulk fields. We denote the fifth coordinate as
$y$ which characterizes $S^{1}$. The orbifold $S^{1}/Z_{2}$ is
constructed by two coordinate identifications $y\rightarrow{-y}$
and $y-\pi{R}\rightarrow{\pi{R-y}}$, with $R$ the $S^{1}$ radius.

The formulation of a gauge theory on $S^{1}/Z_{2}$, requires
$Z_{2}$ to be a symmetry of the lagrangian. The orbifold
identification of coordinates and the requirement of $Z_{2}$
symmetry of the lagrangian, dictates the following field
transformations, under the $Z_{2}$ action:

\begin{equation}
A_{\mu}(x^{\mu},-y)=P{\,}A_{\mu}(x^{\mu},y)P^{\dag}
\end{equation}

\begin{equation}
A_{5}(x^{\mu},-y)=-P{\,}A_{5}(x^{\mu},y)P^{\dag}
\end{equation}

\begin{equation}
\phi(x^{\mu},-y)=\eta T(P){\,}\phi(x^{\mu},y) \label{bos}
\end{equation}

 \begin{equation}
\psi(x^{\mu},-y)=\eta' T(P)\gamma^{5}{\,}\psi(x^{\mu},y)
\label{fer} ,
\end{equation}
where $A_{\mu}$, $A_{5}$, are the $4d$ gauge field and the fifth
component of the gauge field. It is the vacuum expectation value
of the latter that will play the role of the Higgs field. Also
(\ref{bos}) and (\ref{fer}) correspond to the boson and fermion
transformation. $P$ is a suitable $Z_{2}$ representation for gauge
fields components and $T(P)$ is also an appropriate representation
(for example when $\psi$ belongs to the fundamental or the adjoint
representation, $T(P)\psi$ corresponds to $P\psi$ and $P\psi
P^{\dag}$ respectively). Finally $\eta$ and $\eta'$ take the
values $\pm 1$. In the same way, one can substitute $P'$ for $P$
and the transformation $y-\pi{R}\rightarrow{\pi{R-y}}$ and the
same relations hold. In general $P\neq P'$ but the $P=P'$ case
shall be studied in this paper. According to the eigenvalues of
$P,P'$ parities, {\it{i.e.}} ($\pm 1,\pm 1$), the gauge field, for
example, has the following harmonic expansions:
\begin{equation}
A_{\mu}(x^{\mu},y)_{(+,+)}=\frac{1}{\sqrt{2\pi{R}}}\sum^{\infty}_{n=0}A_{\mu}^{(n)}(x^{\mu})_{(+,+)}\cos(\frac{ny}{R})
\end{equation}
\begin{equation}
A_{\mu}(x^{\mu},y)_{(+,-)}=\frac{1}{\sqrt{2\pi{R}}}\sum^{\infty}_{n=0}A_{\mu}^{(n)}(x^{\mu})_{(+,-)}\cos(\frac{(n+\frac{1}{2})y}{R})
\end{equation}
\begin{equation}
A_{\mu}(x^{\mu},y)_{(-,+)}=\frac{1}{\sqrt{2\pi{R}}}\sum^{\infty}_{n=0}A_{\mu}^{(n)}(x^{\mu})_{(-,+)}\sin(\frac{(n+\frac{1}{2})y}{R})
\end{equation}
\begin{equation}
A_{\mu}(x^{\mu},y)_{(-,-)}=\frac{1}{\sqrt{2\pi{R}}}\sum^{\infty}_{n=0}A_{\mu}^{(n)}(x^{\mu})_{(-,-)}\sin(\frac{(n+1)y}{R})
\end{equation}
The expansions of other fields can be done in a similar way. Note
that only the fields having $(+,+)$ parities have zero modes and
thus only these appear as massless particle states at low energy
compared to the compactification scale. The case that zero modes
of $A_{5}$ have branching $(1, 2, 1/2)$ and $(1, 2, -1/2)$ under
$SU(3)_{c}\times SU(2)_{L}\times U(1)_{Y}$ shall be adopted in
this paper (following \cite{hoso2}). These zero modes are regarded
to be the Higgs doublets and this is the essence of gauge-Higgs
unification. It is noticeable and crucial to note that after
radiative corrections, the Higgs mass will be finite, which
follows from $5d$ gauge invariance, that guarantees the
masslessness of $A_{5}$ components. In the case of $N=1$, $5d$
supersymmetry after compactification $A_{5}$ is combined with an
adjoint scalar field. $N=1$, $5d$ supersymmetry corresponds to
$N=2$ supersymmetry in $4d$, where the $5d$ vector multiplet is
composed from a $d=4$ vector multiplet and a $d=4$ chiral
multiplet in the adjoint representation of gauge group ie:
\begin{equation}
V_{5}=(A^{M},\lambda,\lambda',\sigma),
\end{equation}
is decomposed to:
\begin{equation}
V=(A^{\mu},\lambda)
\end{equation}
\begin{equation}
\Sigma(\sigma+i{A_{5}},\lambda'),
\end{equation}
the adjoint vector and adjoint $d=4$ chiral superfields, with
\begin{equation}
V=-\theta\sigma^{\mu}\bar{\theta}A_{\mu}+i\bar{\theta}^{2}\theta
\lambda -i\theta^{2}\bar{\theta} \bar{\lambda}+\frac{1}{2}
\bar{\theta}^{2}\bar{\theta}^{2}D,
\end{equation}
and
\begin{equation}
\Sigma =\frac{1}{\sqrt{2}}(\sigma+iA_{5})+\sqrt{2}\theta
\lambda'+\theta^{2}F.
\end{equation}
The orbifold boundary conditions for the supersymmetric case read:
\begin{equation}
\left(%
\begin{array}{c}
  V(x^{\mu}, -y) \\
 \Sigma{(x^{\mu}, -y)} \\
\end{array}%
\right)=P\left(%
\begin{array}{c}
    V(x^{\mu},y)\\
  -\Sigma{(x^{\mu}, -y)} \\
\end{array}%
\right)P^{\dag},
\end{equation}
and
\begin{equation}
\left(%
\begin{array}{c}
  V(x^{\mu}, \pi{R}-y) \\
 \Sigma{(x^{\mu},\pi{R}-y )} \\
\end{array}%
\right)=P\left(%
\begin{array}{c}
    V(x^{\mu},\pi{R}+y)\\
  -\Sigma{(x^{\mu}, \pi{R}+y)} \\
\end{array}%
\right)P^{\dag}\label{gauge5}
\end{equation}
In this paper we shall use the point of view of \cite{hoso2} and
add extra matter bulk fields (hypermultiplets in terms of $N=2$
supersymmetry in $d=4$) in the adjoint and fundamental
representations of the gauge group. That is, we add $N_{f}$
fundamental hypermultiplet and $N_{\alpha}$ adjoint
hypermultiplet. In $d=4$ the $N=1$, $d=5$ fundamental and adjoint
hypermultiplet, $\Psi$ and $\Psi^{\alpha}$ respectively correspond
to four $N=1$ chiral superfields and anti-chiral {\it{i.e.}}:
\begin{equation}
\Psi=(\phi,\phi^{c\dag},\widetilde{\phi},\widetilde{\phi}^{c\dag}),
\end{equation}
decomposes to $\Phi=(\phi,\widetilde{\phi})$, chiral superfield
and $\Phi^{c}=(\phi^{c},\widetilde{\phi^{c}})$, anti-chiral
superfield. Also
$\Psi^{\alpha}=(\phi^{\alpha},\phi^{\alpha{c\dag}},\widetilde{\phi}^{\alpha},\widetilde{\phi}^{\alpha{c\dag}})$,
decomposes to a chiral superfield
$\Phi=(\phi^{\alpha},\widetilde{\phi}^{\alpha})$ and
$\Phi^{\alpha{c}}=(\phi^{\alpha{c}},\widetilde{\phi}^{\alpha{c}})$,
an anti-chiral adjoint $d=4$ superfield. The orbifold boundary
conditions for the matter superfields are:

\begin{equation}
\left(%
\begin{array}{c}
  \Phi{(x^{\mu}, -y)} \\
 \Phi^{c\dag}{(x^{\mu}, -y)} \\
\end{array}%
\right)=\eta{P}\left(%
\begin{array}{c}
    \Phi{(x^{\mu},y)}\\
   -\Phi^{c\dag}{(x^{\mu},y)}\\
\end{array}%
\right),\label{fund}
\end{equation}

\begin{equation}
\left(%
\begin{array}{c}
  \Phi{(x^{\mu}, \pi{R}-y)} \\
 \Phi^{c\dag}{(x^{\mu}, \pi{R}-y)} \\
\end{array}%
\right)=\eta'{P'}\left(%
\begin{array}{c}
    \Phi{(x^{\mu},\pi{R}+y)}\\
   -\Phi^{c\dag}{(x^{\mu},\pi{R}+y)}\\
\end{array}%
\right),
\end{equation}

\begin{equation}
\left(%
\begin{array}{c}
  \Phi^{\alpha}{(x^{\mu},-y)} \\
 \Phi^{\alpha{c\dag}}{(x^{\mu},-y)} \\
\end{array}%
\right)=\eta{P}\left(%
\begin{array}{c}
    \Phi^{\alpha}{(x^{\mu},y)}\\
   -\Phi^{\alpha{c\dag}}{(x^{\mu},y)}\\
\end{array}%
\right)P^{\dag},
\end{equation}

\begin{equation}
\left(%
\begin{array}{c}
  \Phi^{\alpha}{(x^{\mu}, \pi{R}-y)} \\
 \Phi^{\alpha{c\dag}}{(x^{\mu},\pi{R}-y)} \\
\end{array}%
\right)=\eta'{P'}\left(%
\begin{array}{c}
    \Phi^{\alpha}{(x^{\mu},\pi{R}+y)}\\
   -\Phi^{\alpha{c\dag}}{(x^{\mu},\pi{R}+y)}\\
\end{array}%
\right)P'^{\dag},
\end{equation}
 In the following two sections, two supersymmetric models shall be
treated, $SU(3)_{c}{\times}SU(3)_{w}$ and $SU(5)$, with $P=P'$.

\section*{The $SU(3)_{c}{\times}SU(3)_{w}$ gauge model}

 Consider an $SU(3)_{c}{\times}SU(3)_{w}$ model in $d=5$ with the
representations of $Z_{2}$, $P$ and $P'$, written in the basis of
$SU(3)_{w}$ as:
\begin{equation}
P=P'=\left(%
\begin{array}{ccc}
  1 & 0 & 0 \\
  0 & -1 & 0 \\
  0 & 0 & -1 \\
\end{array}%
\right).
\end{equation}
The components of $V$ and $\Sigma$ (described in the previous
section) transform as:
\begin{equation}
V=\left(%
\begin{array}{ccc}
  (+,+) & (-,-) & (-,-) \\
  (-,-) & (+,+) & (+,+) \\
  (-,-) & (+,+) & (+,+) \\
\end{array}%
\right),
\end{equation}

\begin{equation}
\Sigma=\left(%
\begin{array}{ccc}
  (-,-) & (+,+) & (+,+) \\
  (+,+) & (-,-) & (-,-) \\
  (+,+) & (-,-) & (-,-) \\
\end{array}%
\right)\label{a5},
\end{equation}
under the transformation,
\begin{equation}
\left(%
\begin{array}{c}
  V(x^{\mu}, -y) \\
 \Sigma{(x^{\mu}, -y)} \\
\end{array}%
\right)=P\left(%
\begin{array}{c}
    V(x^{\mu},y)\\
  -\Sigma{(x^{\mu}, -y)} \\
\end{array}%
\right)P^{\dag},
\end{equation}
and
\begin{equation}
\left(%
\begin{array}{c}
  V(x^{\mu}, \pi{R}-y) \\
 \Sigma{(x^{\mu},\pi{R}-y )} \\
\end{array}%
\right)=P\left(%
\begin{array}{c}
    V(x^{\mu},\pi{R}+y)\\
  -\Sigma{(x^{\mu}, \pi{R}+y)} \\
\end{array}%
\right)P^{\dag}.
\end{equation}
 One can see that $SU(3)_{w}$ is broken down to $SU(2)_{L}\times U(1)_{Y}$ by observing the zero modes
$(+,+)$ of the gauge field $V$. Also the zero modes of $\Sigma$
reveal the Higgs doublet structure. It must be noted that in the
above case two Higgs doublets appear and also that $SU(3)_{c}$ is
not broken by orbifold boundary conditions ($P=P'=I$ in the basis
of $SU(3)_{c}$). Now the vev of the scalar part of $\Sigma$ is a
doublet, so the Wilson line degree of freedom utilizing the
residual $SU(2)$ gauge symmetry can be written as \cite{hab1}:
\begin{equation}
\langle \Sigma\rangle=\frac{1}{gR} \sum_{\alpha} \alpha_{\alpha}
\frac{\lambda_{\alpha}}{2}=\left(%
\begin{array}{ccc}
  0 & 0 & \alpha \\
  0 & 0 & 0 \\
  \alpha & 0 & 0 \\
\end{array}%
\right),
\end{equation}
with $g$ the $5d$ gauge coupling. Note the existence of one vev.
The parameter $\alpha$ is related with the Wilson line phases and
determines the further breaking of the gauge symmetry. In this
paper, two symmetry breaking patterns shall be used, determined by
the values of $\alpha$, that is \cite{takenaga}:
\begin{itemize}
\item $\alpha=1$, for which $SU(2)_{L}$ breaks to a $U'(1)$ so the
residual symmetry is $U'(1)\times U(1)_{Y}$

\item $\alpha=0$, for which $SU(2)_{L}$ remains unbroken.
\end{itemize}
These two values correspond to minima of the zero temperature
effective potential \cite{hoso2}.

 Along with the gauge multiplet described above, additional matter fields are added in
the full particle content of the model, specifically $N_{\alpha}$
adjoint and $N_{f}$ fundamental chiral superfields (which are
described below equation (\ref{gauge5})). A further decomposition
is imposed, that is, $N_{\alpha}$ is decomposed to
$N_{\alpha}^{+}$ and $N_{\alpha}^{-}$ and $N_{f}$ to $N_{f}^{+}$
and $N_{f}^{-}$, according to the sign of the product $\eta \eta'$
(see equation (\ref{fund}) and below). $N_{f}^{+}$ and
$N_{\alpha}^{+}$ correspond to $\eta \eta'=1$ and $N_{\alpha}^{-}$
and $N_{f}^{-}$ to $\eta \eta'=-1$ (so fields with $(+,+)$ or
$(-,-)$ parities are counted in the first two).

\subsection*{$SU(3)_{c}{\times}SU(3)_{w}$ at finite temperature}

In this section the finite temperature effective potential
\cite{hab1,takenaga} of each multiplet described in the previous
sections shall be given. The total effective potential, $V_{tot}$,
is equal to the sum of each multiplet contribution. We start with
the gauge ghost gaugino potential $V_{(+)}^{g}$ (the ''+'' sign
corresponds to the $\eta \eta'$ value.):
\begin{align}
V_{(+)}^{g}&=-\frac{4}{2}\int
\frac{dp^{d-2}}{(2\pi)^{d-2}}\frac{T}{R}\sum_{n,m=-\infty}^{\infty}\ln [n^2(\frac{2\pi}{R})^{2}+p^{2}+(2\pi{T})^{2}m^{2}]\notag \\
&+\ln [(n-\alpha)^2(\frac{2\pi}{R})^{2}+p^{2}+(2\pi{T})^{2}m^{2}]\notag \\
&+2\ln [(n-\frac{\alpha}{2})^2(\frac{2\pi}{R})^{2}+p^{2}+(2\pi{T})^{2}m^{2}]\notag \\
&+\frac{4}{2}\int
\frac{dp^{d-2}}{(2\pi)^{d-2}}\frac{T}{R}\sum_{n,m=-\infty}^{\infty}\ln [(n-\alpha+\beta)^2(\frac{2\pi}{R})^{2}+p^{2}+(2\pi{T})^{2}(m+\frac{1}{2})^{2}]\notag \\
&+\ln
[(n+\beta)^2(\frac{2\pi}{R})^{2}+p^{2}+(2\pi{T})^{2}(m+\frac{1}{2})^{2}]\notag \\
&+2\ln
[(n-\frac{\alpha}{2}+\beta)^2(\frac{2\pi}{R})^{2}+p^{2}+(2\pi{T})^{2}(m+\frac{1}{2})^{2}].
\end{align}
where above and in the following the usual convention for finite
temperature field theories was used, that is, periodic bosons and
anti-periodic fermions at finite temperature (a choice consistent
with KMS relations). The Scherk-Schwarz phase $\beta$ has been
added to the gaugino sector \cite{hab1}. The contribution of
$N_{f}^{(+)}$ fermions in the fundamental representation with
$\eta \eta'=1$ is:
\begin{align}
V_{(+)}^{f}&=-\frac{4N_{f}^{(+)}}{2}\int
\frac{dp^{d-2}}{(2\pi)^{d-2}}\frac{T}{R}\sum_{n,m=-\infty}^{\infty}\ln [(n-\frac{\alpha}{2}+\beta)^2(\frac{2\pi}{R})^{2}+p^{2}+(2\pi{T})^{2}m^{2}]\notag \\
&+\ln [(n+\beta)^2(\frac{2\pi}{R})^{2}+p^{2}+(2\pi{T})^{2}m^{2}]\notag \\
&+\frac{4N_{f}^{(+)}}{2}\int
\frac{dp^{d-2}}{(2\pi)^{d-2}}\frac{T}{R}\sum_{n,m=-\infty}^{\infty}\ln [(n+\frac{\alpha}{2})^2(\frac{2\pi}{R})^{2}+p^{2}+(2\pi{T})^{2}(m+\frac{1}{2})^{2}]\notag \\
&+\ln
[n^2(\frac{2\pi}{R})^{2}+p^{2}+(2\pi{T})^{2}(m+\frac{1}{2})^{2}].
\end{align}
and the contribution of $N_{f}^{(-)}$ with $\eta \eta'=-1$:
\begin{align}
V_{(-)}^{f}&=-\frac{4N_{f}^{(-)}}{2}\int
\frac{dp^{d-2}}{(2\pi)^{d-2}}\frac{T}{R}\sum_{n,m=-\infty}^{\infty}\ln [(n-\frac{\alpha}{2}+\frac{1}{2}+\beta)^2(\frac{2\pi}{R})^{2}+p^{2}+(2\pi{T})^{2}m^{2}]\notag \\
&+\ln [(n+\frac{1}{2}+\beta)^2(\frac{2\pi}{R})^{2}+p^{2}+(2\pi{T})^{2}m^{2}]\notag \\
&+\frac{4N_{f}^{(-)}}{2}\int
\frac{dp^{d-2}}{(2\pi)^{d-2}}\frac{T}{R}\sum_{n,m=-\infty}^{\infty}\ln [(n-\frac{\alpha}{2}+\frac{1}{2})^2(\frac{2\pi}{R})^{2}+p^{2}+(2\pi{T})^{2}(m+\frac{1}{2})^{2}]\notag \\
&+\ln
[(n+\frac{1}{2})^2(\frac{2\pi}{R})^{2}+p^{2}+(2\pi{T})^{2}(m+\frac{1}{2})^{2}].
\end{align}
Notice the difference in the two contributions coming from the
orbifold boundary conditions and the different Wilson line phase
eigenvalue. The effective potential for $N^{(+)}_{\alpha}$ adjoint
matter field chiral multiplet, with $\eta \eta'=1$ is:
\begin{align}
V_{(+)}^{\alpha}&=-\frac{4N^{(+)}_{\alpha}}{2}\int
\frac{dp^{d-2}}{(2\pi)^{d-2}}\frac{T}{R}\sum_{n,m=-\infty}^{\infty}\ln [n^2(\frac{2\pi}{R})^{2}+p^{2}+(2\pi{T})^{2}m^{2}]\notag \\
&+\ln [(n-\alpha)^2(\frac{2\pi}{R})^{2}+p^{2}+(2\pi{T})^{2}m^{2}]\notag \\
&+2\ln [(n-\frac{\alpha}{2})^2(\frac{2\pi}{R})^{2}+p^{2}+(2\pi{T})^{2}m^{2}]\notag \\
&+\frac{4N^{(+)}_{\alpha}}{2}\int
\frac{dp^{d-2}}{(2\pi)^{d-2}}\frac{T}{R}\sum_{n,m=-\infty}^{\infty}\ln [(n-\alpha+\beta)^2(\frac{2\pi}{R})^{2}+p^{2}+(2\pi{T})^{2}(m+\frac{1}{2})^{2}]\notag \\
&+\ln
[(n+\beta)^2(\frac{2\pi}{R})^{2}+p^{2}+(2\pi{T})^{2}(m+\frac{1}{2})^{2}]\notag \\
&+2\ln
[(n-\frac{\alpha}{2}+\beta)^2(\frac{2\pi}{R})^{2}+p^{2}+(2\pi{T})^{2}(m+\frac{1}{2})^{2}].
\end{align}
Note that for $N^{(+)}_{\alpha}=1$ the adjoint chiral multiplet
together with the gauge vector multiplet correspond to $N=2$,
$d=4$ supersymmetry. Finally the contribution of
$N^{(-)}_{\alpha}$ adjoint chiral superfield with $\eta \eta'=-1$
is:
\begin{align}
V_{(-)}^{\alpha}&=\frac{4N^{(-)}_{\alpha}}{2}\int
\frac{dp^{d-2}}{(2\pi)^{d-2}}\frac{T}{R}\sum_{n,m=-\infty}^{\infty}\ln [(n+\frac{1}{2})^2(\frac{2\pi}{R})^{2}+p^{2}+(2\pi{T})^{2}(m+\frac{1}{2})^{2}]\notag \\
&+\ln [(n-\alpha+\frac{1}{2})^2(\frac{2\pi}{R})^{2}+p^{2}+(2\pi{T})^{2}(m+\frac{1}{2})^{2}]\notag \\
&+2\ln [(n-\frac{\alpha}{2}+\frac{1}{2})^2(\frac{2\pi}{R})^{2}+p^{2}+(2\pi{T})^{2}(m+\frac{1}{2})^{2}]\notag \\
&-\frac{4N^{(-)}_{\alpha}}{2}\int
\frac{dp^{d-2}}{(2\pi)^{d-2}}\frac{T}{R}\sum_{n,m=-\infty}^{\infty}\ln [(n-\alpha+\beta+\frac{1}{2})^2(\frac{2\pi}{R})^{2}+p^{2}+(2\pi{T})^{2}m^{2}]\notag \\
&+\ln
[(n+\beta+\frac{1}{2})^2(\frac{2\pi}{R})^{2}+p^{2}+(2\pi{T})^{2}m^{2}]\notag \\
&+2\ln
[(n-\frac{\alpha}{2}+\beta+\frac{1}{2})^2(\frac{2\pi}{R})^{2}+p^{2}+(2\pi{T})^{2}m^{2}].
\end{align}
The total effective potential for the above particle ensemble at
finite temperature is written:
\begin{equation}
V_{tot}=V_{(+)}^{g}+V_{(+)}^{f}+V_{(+)}^{\alpha}+V_{(-)}^{\alpha}+V_{(-)}^{f}
\end{equation}

\subsection*{The $\alpha=1$, $\beta=\frac{1}{2}$ case}

The case $\alpha=1$, $\beta=\frac{1}{2}$ is very interesting since
the remaining $SU(2)$ gauge symmetry that ''survived'' the
orbifolding breaks dynamically to a $U'(1)$
\cite{hoso1,hab1,takenaga} through the Wilson line phase at the
minimum $\alpha=1$. The contributions for the fields become:
\begin{align}
V_{(+)}^{g}&=-\frac{4}{2}\int
\frac{dp^{d-2}}{(2\pi)^{d-2}}\frac{T}{R}\sum_{n,m=-\infty}^{\infty}2\ln [n^2(\frac{2\pi}{R})^{2}+p^{2}+(2\pi{T})^{2}m^{2}]\notag \\
&+2\ln [(n+\frac{1}{2})^2(\frac{2\pi}{R})^{2}+p^{2}+(2\pi{T})^{2}m^{2}]\notag \\
&+\frac{4}{2}\int
\frac{dp^{d-2}}{(2\pi)^{d-2}}\frac{T}{R}\sum_{n,m=-\infty}^{\infty}2\ln [(n+\frac{1}{2})^2(\frac{2\pi}{R})^{2}+p^{2}+(2\pi{T})^{2}(m+\frac{1}{2})^{2}]\notag \\
&+2\ln
[n^2(\frac{2\pi}{R})^{2}+p^{2}+(2\pi{T})^{2}(m+\frac{1}{2})^{2}],
\end{align}
for the gauge, ghost and gaugino. For $N_{f}^{(-)}$ fundamental
fermions with $\eta \eta'=-1$:
\begin{align}
V_{(-)}^{f}&=-\frac{4N_{f}^{(-)}}{2}\int
\frac{dp^{d-2}}{(2\pi)^{d-2}}\frac{T}{R}\sum_{n,m=-\infty}^{\infty}\ln [(n+\frac{1}{2})^2(\frac{2\pi}{R})^{2}+p^{2}+(2\pi{T})^{2}m^{2}]\notag \\
&+\ln [n^2(\frac{2\pi}{R})^{2}+p^{2}+(2\pi{T})^{2}m^{2}]\notag \\
&+\frac{4N_{f}^{(-)}}{2}\int
\frac{dp^{d-2}}{(2\pi)^{d-2}}\frac{T}{R}\sum_{n,m=-\infty}^{\infty}\ln [n^2(\frac{2\pi}{R})^{2}+p^{2}+(2\pi{T})^{2}(m+\frac{1}{2})^{2}]\notag \\
&+\ln
[(n+\frac{1}{2})^2(\frac{2\pi}{R})^{2}+p^{2}+(2\pi{T})^{2}(m+\frac{1}{2})^{2}],
\end{align}
and for $N_{f}^{(+)}$ fundamental fermions with $\eta \eta'=1$:
\begin{align}
V_{(+)}^{f}&=-\frac{4N_{f}^{(+)}}{2}\int
\frac{dp^{d-2}}{(2\pi)^{d-2}}\frac{T}{R}\sum_{n,m=-\infty}^{\infty}\ln [n^2(\frac{2\pi}{R})^{2}+p^{2}+(2\pi{T})^{2}m^{2}]\notag \\
&+\ln [(n+\frac{1}{2})^2(\frac{2\pi}{R})^{2}+p^{2}+(2\pi{T})^{2}m^{2}]\notag \\
&+\frac{4N_{f}^{(+)}}{2}\int
\frac{dp^{d-2}}{(2\pi)^{d-2}}\frac{T}{R}\sum_{n,m=-\infty}^{\infty}\ln [(n+\frac{1}{2})^2(\frac{2\pi}{R})^{2}+p^{2}+(2\pi{T})^{2}(m+\frac{1}{2})^{2}]\notag \\
&+\ln
[n^2(\frac{2\pi}{R})^{2}+p^{2}+(2\pi{T})^{2}(m+\frac{1}{2})^{2}].
\end{align}
Finally the potential for $N^{(+)}_{\alpha}$ adjoint chiral
superfields, with $\eta \eta'=1$
\begin{align}
V_{(+)}^{\alpha}&=-\frac{4N^{(+)}_{\alpha}}{2}\int
\frac{dp^{d-2}}{(2\pi)^{d-2}}\frac{T}{R}\sum_{n,m=-\infty}^{\infty}2\ln [n^2(\frac{2\pi}{R})^{2}+p^{2}+(2\pi{T})^{2}m^{2}]\notag \\
&+2\ln [(n+\frac{1}{2})^2(\frac{2\pi}{R})^{2}+p^{2}+(2\pi{T})^{2}m^{2}]\notag \\
&+\frac{4N^{(+)}_{\alpha}}{2}\int
\frac{dp^{d-2}}{(2\pi)^{d-2}}\frac{T}{R}\sum_{n,m=-\infty}^{\infty}2\ln [(n+\frac{1}{2})^2(\frac{2\pi}{R})^{2}+p^{2}+(2\pi{T})^{2}(m+\frac{1}{2})^{2}]\notag \\
&+2\ln
[n^2(\frac{2\pi}{R})^{2}+p^{2}+(2\pi{T})^{2}(m+\frac{1}{2})^{2}],
\end{align}
and for $N^{(-)}_{\alpha}$ adjoint with $\eta \eta'=-1$
\begin{align}
V_{(-)}^{\alpha}&=\frac{4N^{(-)}_{\alpha}}{2}\int
\frac{dp^{d-2}}{(2\pi)^{d-2}}\frac{T}{R}\sum_{n,m=-\infty}^{\infty}2\ln [(n+\frac{1}{2})^2(\frac{2\pi}{R})^{2}+p^{2}+(2\pi{T})^{2}(m+\frac{1}{2})^{2}]\notag \\
&+2\ln [n^2(\frac{2\pi}{R})^{2}+p^{2}+(2\pi{T})^{2}(m+\frac{1}{2})^{2}]\notag \\
&-\frac{4N^{(-)}_{\alpha}}{2}\int
\frac{dp^{d-2}}{(2\pi)^{d-2}}\frac{T}{R}\sum_{n,m=-\infty}^{\infty}2\ln [n^2(\frac{2\pi}{R})^{2}+p^{2}+(2\pi{T})^{2}m^{2}]\notag \\
&+2\ln
[(n+\frac{1}{2})^2(\frac{2\pi}{R})^{2}+p^{2}+(2\pi{T})^{2}m^{2}].
\end{align}
Again, the total effective potential $V_{tot}$ is:
\begin{equation}
V_{tot}=V_{(+)}^{g}+V_{(+)}^{f}+V_{(+)}^{\alpha}+V_{(-)}^{\alpha}+V_{(-)}^{f}.
\end{equation}
Upon using,
\begin{equation}
\int \frac{dk^{d}}{(2\pi )^{d}}\ln (k^{2}+a^{2})=-\frac{\Gamma (-\frac{d}{2})%
}{(4\pi )^{\frac{d}{2}}}a^{d},  \label{dimreg}
\end{equation}
the total contribution of $N_{\alpha}^{+}$, $N_{\alpha}^{-}$,
$N_{f}^{+}$, $N_{f}^{-}$ and of the gauge-gaugino, to the
effective potential becomes,
\begin{align}
V_{tot}
=&\frac{2T}{R}\Big{(}N_{f}^{-}+N_{f}^{+}+2+2N_{\alpha}^{-}+2N_{\alpha}^{+}\Big{)}
\frac{\Gamma (\frac{2-d}{2})}{(4\pi
)^{\frac{d-2}{2}}}\times\\\notag &{\bigg (}\sum_{n,m=-\infty
}^{\infty } {\Big (}(2\pi
T)^{2}m^{2}+(n+\frac{1}{2})^{2}\big{(}\frac{2\pi}{R}{\big{)}}^{2}{\Big
)}^{ \frac{d-2}{2}} \\\notag &-\sum_{n,m=-\infty }^{\infty }
{\Big(}((2\pi
T)^{2}(m+\frac{1}{2})^{2}+n^{2}{\big{(}}\frac{2\pi}{R}{\big{)}}^{2}{\Big)}^{^{\frac{d-2}{2}}}{\bigg
)}\\\notag
&+\frac{2T}{R}\Big{(}N_{f}^{-}+N_{f}^{+}+2+2N_{\alpha}^{-}+2N_{\alpha}^{+}\Big{)}
\frac{\Gamma (\frac{2-d}{2})}{(4\pi
)^{\frac{d-2}{2}}}\times\\\notag &{\bigg (}\sum_{n,m=-\infty
}^{\infty } {\Big (}(2\pi
T)^{2}m^{2}+n^{2}\big{(}\frac{2\pi}{R}{\big{)}}^{2}{\Big )}^{
\frac{d-2}{2}} \\\notag &-\sum_{n,m=-\infty }^{\infty }
{\Big(}((2\pi
T)^{2}(m+\frac{1}{2})^{2}+(n+\frac{1}{2})^{2}{\big{(}}\frac{2\pi}{R}{\big{)}}^{2}{\Big)}^{^{\frac{d-2}{2}}}{\bigg
)}
\end{align}
By introducing the dimensionless parameter $\xi =RT$, we obtain,
\begin{align}
V_{tot}
=&\frac{2T}{R}\Big{(}N_{f}^{-}+N_{f}^{+}+2+2N_{\alpha}^{-}+2N_{\alpha}^{+}\Big{)}
{\big{(}}\frac{2\pi}{R}{\big{)}}^{d-2}\frac{\Gamma
(\frac{2-d}{2})}{(4\pi )^{\frac{d-2}{2}}}\times\\\notag &{\bigg
(}\sum_{n,m=-\infty }^{\infty } {\Big
(}{\xi}^{2}m^{2}+(n+\frac{1}{2})^{2}{\Big )}^{
\frac{d-2}{2}}-\sum_{n,m=-\infty }^{\infty }
{\Big(}{\xi}^{2}(m+\frac{1}{2})^{2}+n^{2}{\Big)}^{^{\frac{d-2}{2}}}{\bigg
)}\\\notag
&+\frac{2T}{R}\Big{(}N_{f}^{-}+N_{f}^{+}+2+2N_{\alpha}^{-}+2N_{\alpha}^{+}\Big{)}
{\big{(}}\frac{2\pi}{R}{\big{)}}^{d-2}\frac{\Gamma
(\frac{2-d}{2})}{(4\pi )^{\frac{d-2}{2}}}\times\\\notag &{\bigg
(}\sum_{n,m=-\infty }^{\infty } {\Big (}{\xi}^{2}m^{2}+n^{2}{\Big
)}^{ \frac{d-2}{2}}-\sum_{n,m=-\infty }^{\infty }
{\Big(}{\xi}^{2}(m+\frac{1}{2})^{2}+(n+\frac{1}{2})^{2}{\Big)}^{^{\frac{d-2}{2}}}{\bigg
)},
\end{align}
and with the aid of two dimensional Epstein zeta
\cite{elizalde,goncharov}
\begin{equation*}
Z_{2}\left\vert
\begin{array}{cc}
g_{1} & g_{2} \\
h_{1} & h_{2}%
\end{array}%
\right\vert (a,a_{1},a_{2})=\sum\limits_{n,m=-\infty }^{\infty}
(a_{1}(n+g_{1})^{2}+a_{2}(m+g_{2})^{2})^{-a}\times \exp [2\pi
i(nh_{1}+mh_{2})],
\end{equation*}%
we obtain,
\begin{align}
V_{tot}
=&\frac{2T}{R}\Big{(}N_{f}^{-}+N_{f}^{+}+2+2N_{\alpha}^{-}+2N_{\alpha}^{+}\Big{)}
{\big{(}}\frac{2\pi}{R}{\big{)}}^{d-2}\frac{\Gamma
(\frac{2-d}{2})}{(4\pi )^{\frac{d-2}{2}}}\times\\\notag &{\bigg
(}Z_{2}\left\vert
\begin{array}{cc}
0 & \frac{1}{2} \\
0 & 0
\end{array}
\right\vert (\frac{2-d}{2},\xi ^{2},1)-Z_{2}\left\vert
\begin{array}{cc}
\frac{1}{2} & 0 \\
0 & 0%
\end{array}
\right\vert (\frac{2-d}{2},\xi ^{2},1){\bigg )}\\\notag
&+\frac{2T}{R}\Big{(}N_{f}^{-}+N_{f}^{+}+2+2N_{\alpha}^{-}+2N_{\alpha}^{+}\Big{)}
{\big{(}}\frac{2\pi}{R}{\big{)}}^{d-2}\frac{\Gamma
(\frac{2-d}{2})}{(4\pi )^{\frac{d-2}{2}}}\times\\\notag &{\bigg
(}Z_{2}\left\vert
\begin{array}{cc}
0 & 0 \\
0 & 0
\end{array}
\right\vert (\frac{2-d}{2},\xi ^{2},1)-Z_{2}\left\vert
\begin{array}{cc}
\frac{1}{2} & \frac{1}{2} \\
0 & 0%
\end{array}
\right\vert (\frac{2-d}{2},\xi ^{2},1){\bigg )}. \notag
\end{align}
To extend analytically the two dimensional Epstein zeta, to values
$Rea<1$, we use the functional equation,
\begin{align}
&Z_{2}\left\vert
\begin{array}{cc}
g_{1} & g_{2} \\
h_{1} & h_{2}%
\end{array}%
\right\vert (a,a_{1},a_{2}) =  \label{epstein} \\
&(a_{1}a_{2})^{-\frac{1}{2}}\pi ^{2a-1}\frac{\Gamma (1-a)}{\Gamma (a)}%
\times \exp [-2\pi i(g_{1}h_{1}+g_{2}h_{2})]  \notag \\
&\times Z_{2}\left\vert
\begin{array}{cc}
h_{1} & h_{2} \\
-g_{1} & -g_{2}%
\end{array}%
\right\vert (1-a,\frac{1}{a_{1}},\frac{1}{a_{2}}),\notag
\end{align}
and $V_{tot}$ after some algebra reads,
\begin{align}
V_{tot}
=&2{\pi}^{-\frac{d}{2}}\Big{(}N_{f}^{-}+N_{f}^{+}+2+2N_{\alpha}^{-}+2N_{\alpha}^{+}\Big{)}
{\xi}^{d}\frac{\Gamma (\frac{d}{2})}{R^{d}}\times\\\notag &{\bigg
(}Z_{2}\left\vert
\begin{array}{cc}
0 & 0 \\
0 & -\frac{1}{2}
\end{array}
\right\vert (\frac{d}{2},1,\xi ^{2})-Z_{2}\left\vert
\begin{array}{cc}
0 & 0 \\
-\frac{1}{2} & 0
\end{array}
\right\vert (\frac{d}{2},1,\xi ^{2}){\bigg )}\\\notag
&2{\pi}^{-\frac{d}{2}}\Big{(}N_{f}^{-}+N_{f}^{+}+2+2N_{\alpha}^{-}+2N_{\alpha}^{+}\Big{)}
{\xi}^{d}\frac{\Gamma (\frac{d}{2})}{R^{d}}\times\\\notag &{\bigg
(}Z_{2}\left\vert
\begin{array}{cc}
0 & 0 \\
0 & 0
\end{array}
\right\vert(\frac{d}{2},1,\xi ^{2})-Z_{2}\left\vert
\begin{array}{cc}
0 & 0 \\
-\frac{1}{2} & -\frac{1}{2}
\end{array}
\right\vert (\frac{d}{2},1,\xi ^{2}){\bigg )}.  \notag
\end{align}
Now set,
\begin{equation}
f(\xi)={\xi}^{d}{\bigg (}Z_{2}\left\vert
\begin{array}{cc}
0 & 0 \\
0 & -\frac{1}{2}
\end{array}
\right\vert (\frac{d}{2},1,\xi ^{2})-Z_{2}\left\vert
\begin{array}{cc}
0 & 0 \\
-\frac{1}{2} & 0
\end{array}
\right\vert (\frac{d}{2},1,\xi ^{2}){\bigg )}\label{f1},
\end{equation}
and
\begin{equation}
g(\xi)={\xi}^{d}{\bigg (}Z_{2}\left\vert
\begin{array}{cc}
0 & 0 \\
0 & 0
\end{array}
\right\vert (\frac{d}{2},1,\xi ^{2})-Z_{2}\left\vert
\begin{array}{cc}
0 & 0 \\
-\frac{1}{2} & -\frac{1}{2}
\end{array}
\right\vert (\frac{d}{2},1,\xi ^{2}){\bigg )}.
\end{equation}
One can see that,
\begin{eqnarray}
f(\frac{1}{\xi })&=&\frac{1}{\xi ^{d}} {\bigg (}Z_{2}\left\vert
\begin{array}{cc}
0 & 0 \\
0 & -\frac{1}{2}%
\end{array}%
\right\vert (\frac{d}{2},1,\frac{1}{\xi ^{2}})-Z_{2}\left\vert
\begin{array}{cc}
0 & 0 \\
-\frac{1}{2} & 0%
\end{array}%
\right\vert (\frac{d}{2},1,\frac{1}{\xi ^{2}}){\bigg )}\label{f2},
\end{eqnarray}%
or equivalently,
\begin{equation}
f(\frac{1}{\xi })=  Z_{2}\left\vert
\begin{array}{cc}
0 & 0 \\
0 & -\frac{1}{2}%
\end{array}%
\right\vert (\frac{d}{2},\xi ^{2},1)-Z_{2}\left\vert
\begin{array}{cc}
0 & 0 \\
-\frac{1}{2} & 0%
\end{array}%
\right\vert (\frac{d}{2},\xi ^{2},1).  \label{inverse}
\end{equation}
From the last expression we easily obtain:%
\begin{align}
&Z_{2}\left\vert
\begin{array}{cc}
0 & 0 \\
0 & -\frac{1}{2}%
\end{array}%
\right\vert (\frac{d}{2},\xi ^{2},1)-Z_{2}\left\vert
\begin{array}{cc}
0 & 0 \\
-\frac{1}{2} & 0%
\end{array}%
\right\vert (\frac{d}{2},\xi ^{2},1) =  \label{inverse2} \\
&-{\bigg (}Z_{2}\left\vert
\begin{array}{cc}
0 & 0 \\
0 & -\frac{1}{2}%
\end{array}%
\right\vert (\frac{d}{2},1,\xi ^{2})-Z_{2}\left\vert
\begin{array}{cc}
0 & 0 \\
-\frac{1}{2} & 0%
\end{array}%
\right\vert (\frac{d}{2},1,\xi ^{2}){\bigg )}.  \notag
\end{align}%
Then combining equations (\ref{f1}), (\ref{f2}), (\ref{inverse2}),
we obtain:
\begin{equation}
f(\xi )=-\xi ^{d}f(\frac{1}{\xi })  \label{antiduality}
\end{equation}
and in the same manner, for $g(\xi )$ we get,
\begin{equation}
g(\xi )=\xi ^{d}f(\frac{1}{\xi })  \label{duality}
\end{equation}
So the total effective potential $V_{tot}$ is written:
\begin{equation}
C_{1}V_{tot}=f(\xi)+g(\xi)
\end{equation}
with,
\begin{equation}
C_{1}=\frac{R^{d}}{2{\pi}^{-\frac{d}{2}}\Big{(}N_{f}^{-}+N_{f}^{+}+2+2N_{\alpha}^{-}+2N_{\alpha}^{+}\Big{)}\Gamma(\frac{d}{2})}.
\end{equation}

 Notice that the temperature inversion symmetry is lost in the full effective
potential and there exist a totally symmetric part $g(\xi )$ and
totally anti-symmetric part $f(\xi )$ in the effective potential.

\subsection*{The $\alpha=0$, $\beta=\frac{1}{2}$ case}

Now the case $\alpha=0$, $\beta=\frac{1}{2}$ follows. However this
case is less interesting from a phenomenological point of view
because $SU(2)_{L}$ remains unbroken. Despite that, we quote the
results to see whether temperature inversion symmetry holds.
Following the analysis of the previous section the total effective
potential is calculated to be:
\begin{align}
V_{tot} =&2{\pi}^{-\frac{d}{2}}\Big{(}2N_{f}^{+}\Big{)}
{\xi}^{d}\frac{\Gamma (\frac{d}{2})}{R^{d}}\times\\\notag &{\bigg
(}Z_{2}\left\vert
\begin{array}{cc}
0 & 0 \\
0 & -\frac{1}{2}
\end{array}
\right\vert (\frac{d}{2},1,\xi ^{2})-Z_{2}\left\vert
\begin{array}{cc}
0 & 0 \\
-\frac{1}{2} & 0
\end{array}
\right\vert (\frac{d}{2},1,\xi ^{2}){\bigg )}\\\notag
&+2{\pi}^{-\frac{d}{2}}\Big{(}2N_{f}^{-}+4+4N_{\alpha}^{-}+4N_{\alpha}^{+}\Big{)}
{\xi}^{d}\frac{\Gamma (\frac{d}{2})}{R^{d}}\times\\\notag &{\bigg
(}Z_{2}\left\vert
\begin{array}{cc}
0 & 0 \\
0 & 0
\end{array}
\right\vert(\frac{d}{2},1,\xi ^{2})-Z_{2}\left\vert
\begin{array}{cc}
0 & 0 \\
-\frac{1}{2} & -\frac{1}{2}
\end{array}
\right\vert (\frac{d}{2},1,\xi ^{2}){\bigg )}.  \notag
\end{align}
It is easily proved that,
\begin{align}
V_{A} =&2{\pi}^{-\frac{d}{2}}\Big{(}2N_{f}^{+}\Big{)}
{\xi}^{d}\frac{\Gamma (\frac{d}{2})}{R^{d}}\times\\\notag &{\bigg
(}Z_{2}\left\vert
\begin{array}{cc}
0 & 0 \\
0 & -\frac{1}{2}
\end{array}
\right\vert (\frac{d}{2},1,\xi ^{2})-Z_{2}\left\vert
\begin{array}{cc}
0 & 0 \\
-\frac{1}{2} & 0
\end{array}
\right\vert (\frac{d}{2},1,\xi ^{2}){\bigg )},\\\notag
\end{align}
is antisymmetric under temperature inversion symmetry while,
\begin{align}
V_{S}=&+2{\pi}^{-\frac{d}{2}}\Big{(}2N_{f}^{-}+4+4N_{\alpha}^{-}+4N_{\alpha}^{+}\Big{)}
{\xi}^{d}\frac{\Gamma (\frac{d}{2})}{R^{d}}\times\\\notag &{\bigg
(}Z_{2}\left\vert
\begin{array}{cc}
0 & 0 \\
0 & 0
\end{array}
\right\vert(\frac{d}{2},1,\xi ^{2})-Z_{2}\left\vert
\begin{array}{cc}
0 & 0 \\
-\frac{1}{2} & -\frac{1}{2}
\end{array}
\right\vert (\frac{d}{2},1,\xi ^{2}){\bigg )},  \notag
\end{align}
is symmetric under temperature inversion. Comparing the $\alpha=0$
and $\alpha=1$ case, it can be seen that in the first case only
the fundamental fermions with $(+,+)$ or $(-,-)$ parities break
the temperature inversion symmetry while, in the second case, all
particle species contain a symmetric and antisymmetric part. A
simple analysis can show that due to orbifold boundary conditions
between the fundamental fermion multiplet components, the
temperature inversion symmetry cannot be a symmetry of the
effective potential, if $(+,+)$ or $(-,-)$ fermions contribute to
the effective potential. This result is of particular importance
since in some extra dimensions models, all the matter (flavor)
fermions (leptons and quarks) are localized on the orbifold fixed
points and do not have KK excitations nor zero modes. The result
implies that in models with localized fundamental fermions on the
orbifold fixed points, temperature inversion symmetry will be a
symmetry of the bulk effective potential (using the conventions of
the previous analysis, that is, $\alpha=0$ and
$\beta=\frac{1}{2}$).

Finally it must be mentioned that the choices for the $\alpha$ and
$\beta$ parameters where the most obvious ones, that is,
$\alpha=0,1$ (for which values $V_{tot}$ has local minima and
gauge symmetry breaks ($\alpha=1$) or not ($\alpha=0$), see
\cite{hoso2}) and $\beta=\frac{1}{2}$ (which gives total
periodicity or anti-periodicity under $y\rightarrow y+2\pi R$
\cite{qui3}). Also care has been made to study phenomenologically
viable cases (for example giving the gauge bosons a Scherk-Schwarz
phase, would be unacceptable phenomenologically).

Of course there is always one question in gauge-Higgs unification
models, which value of $\alpha$ and $\beta$ give correct
electroweak breaking. The values used in this paper don't give
correct electroweak symmetry breaking. Numerical analysis made in
\cite{hoso2,takenaga} give reliable results. Our investigation
involved the most obvious values for gauge symmetry breaking (that
are minima of the effective potential) and periodicity
anti-periodicity under $y\rightarrow y+2\pi R$ and supersymmetry
breaking.

\bigskip
\section*{Conclusions}
Dualities play an important role especially in effective field
theories with predictions that are energetically higher than the
current experimental bound or perturbatively unreachable. In this
paper, we studied a kind of duality the temperature inversion
symmetry \cite{wotzasek} $R\rightarrow \frac{1}{T}$, which
connects the Boltzmann free energy, with the zero temperature
vacuum energy of a system. This was done for the supersymmetric
$d=5$, $SU(3)_{w}\times SU(3)_{c}$ gauge-Higgs unification model.
The value of the Scherk-Schwarz breaking phase used was
$\beta=\frac{1}{2}$ and that of the Wilson line phase was
$\alpha=0,1$. The value of $\beta$ corresponds to periodicity or
anti-periodicity of the fields under $y\rightarrow y+2\pi R$ in
the first case \cite{qui2} and the values $\alpha=0,1$ are the
minimum of the zero temperature effective potential \cite{hoso2}.
For $\alpha=1$, it is found that, irrespectively what the number
of hypermultiplets is, the total effective potential $V_{tot}$ has
an anti-symmetric part under $R\rightarrow \frac{1}{T}$ (or
equivalently $\xi\rightarrow \frac{1}{\xi}$ with $\xi=RT$)
expressed in terms of $f(\xi)$:
\begin{equation}
f(\xi )=-\xi ^{d}f(\frac{1}{\xi }),
\end{equation}
and a symmetric part $g(\xi)$:
\begin{equation}
g(\xi )=\xi ^{d}f(\frac{1}{\xi }),
\end{equation}
with
\begin{equation}
C_{1}V_{tot}=f(\xi)+g(\xi),
\end{equation}
and $C_{1}$ a constant. In the above $\xi=RT$, where $R$ and $T$
are the compactification radius and the temperature of the system,
respectively. For the case $\alpha=1$ the remaining, after the
orbifold compactification, $SU(2)_{L}$ gauge symmetry, breaks to a
$U'(1)$. The case $\alpha=0$ corresponds to unbroken $SU(2)_{L}$
gauge symmetry. In this case an interesting result has been found.
It was shown in the previous sections, that the contribution of
fermions in the fundamental representation of the gauge group,
with $(+,+)$ or $(-,-)$ $Z_{2}$ parities, is the only contribution
to the effective potential that spoils temperature inversion
symmetry. Thus in models with orbifold fixed point localized
fundamental fermions (without bulk fundamental fermions), the
temperature inversion symmetry holds for the bulk effective
potential. This has been examined extensively and for every case
(the Scherk-Schwarz parameter was added in the fundamental
s-particles), but the result remained the same.

Finally we must note that in order to get correct electroweak
breaking in these models, $\alpha$ must take fractional values.
For an excellent treatment of these issues see
\cite{hoso2,takenaga}. However for that values the symmetry would
be completely destroyed.

It seems that orbifolding affects the temperature inversion
symmetry of the effective potential, corresponding to a field
theory system, quantized on an orbifold. The same is also expected
for any ensemble quantized on a product manifold with twisted
boundary conditions. Thus the details of the boundary conditions
determine whether the symmetry holds or not.

\bigskip

\section*{Acknowledgements}
V.K.O would like to thank N. Trakas, S. Massen, Ch. Moustakidis
and D.Iakovidou

\end{document}